\def\nucnode{{\mathcal N}}
\def\startnode{{\mathcal S}}
\def\endnode{{\mathcal E}}

\def\s3{\sqrt{3}}

\def\c0{{\underline 0}}

\def\a{\alpha}

\def\s{\sigma}

\def\Q3{{\bf Q}3} 

\def\Q{{\bf Q}}

\def\a{\alpha}

\def\A{{\bf A}}
\def\B{{\bf B}}

\documentclass[12pt]{article}
\usepackage{latexsym}
\usepackage{graphicx}
\begin{document}
\title{Probabilistic pairwise sequence alignment}
\author{Lawren Smithline}
\date{\today}
\maketitle
\abstract{
We describe an new algorithm for visualizing an alignment of
biological sequences according to a probabilistic model of
evolution.  The resulting data array is readily
interpreted 
by the human eye and amenable to digital image techniques.

We present examples using mRNA sequences from mouse and rat:
three cytochromes and two zinc finger proteins.
The underlying evolutionary model is derived from one proposed
by Thorne, Kishino, and Felsenstein 
and improved by Hein and others. 
The demonstration
implementation aligns two sequences using time and memory quadratic
in the mean sequence length.

The algorithm is extensible, after Lunter, Mikl\'os, Song and Hein
to multiple sequences.  We mention a basic method to reduce time
and memory demands.
\\
}
\section{Introduction}
The problems of inferring common ancestors and 
determining the evolution of observed biological sequences
have received much attention.  The
challenge becomes greater with the interest in mining genome scale
sequences for meaning.  The most popular tools produce mountains
of data which are hard to assimilate.

Current solution strategies for sequence alignment
begin by postulating some model of evolution
in order to score proposed alignments.  This transforms the task into
development of good score functions and
the search for alignments with high score.

The BLAST family of tools uses a hash-and-extend method to find regions
of the test sequences which have high similarity, measured by its
score model.  A hash-and-extend method searces for identical short
subsequences by a very fast hash table search and extends these short
exact matches to longer not quite exact matches.

Dynamic programming tools, including CLUSTAL, use a score array to
find the best scoring global alignment.  Needleman-Wunsh and Smith-Waterman
are classic algorithms for the two sequence global alignment problem.

Both of these approaches are useful for aligning sequences when
the true picture is a one-to-one correspondence, possibly with simple
deletions or insertions.

BLAST picks out one-to-many similarities.
CLUSTAL finds a one-to-one alignment of whole sequences or alignment
profiles.
Neither is optimal for the other task.

Both approaches are fooled by sequences which have internal
near repeats.  This can happen when there are duplications
from part of an ancestor sequence in each of its descendants,
When the duplications are not identical for each descendant or
duplications have independently mutated, there are many regions in
one sequence similar to many regions in another.
CLUSTAL chooses one path, which maximizes global score, but
may have a nonsense agglomeration of gaps.  BLAST typically
reports every pair of similar subsequences.  It is an easy {\em in silico}
demonstration that BLAST can miss similiarities entirely
if mutations are peppered just frequently enough.
\\

Our goal is to develop an alignment method which finds
good global alignments and does not cut out those alignments
which score well but not highest.

Merits of our score function are: it is meaningful as a probability;
it describes locally the value of an alignment at a point; and
it is comparable within an alignment and across alignments.
These features permit detection of repeats and duplications.

We concentrate on developing technique for pairwise alignment in a way
which is extensible to small numbers of sequences.
A larger goal, not yet realized, is multiple genomic alignments.
Toward this end, we suggest a conceptually easy way
to avoid computing on regions which score so poorly,
that they are unlikely to contribute to any of the good alignments. 
It is mentioned after the algorithm description for
clarity of the exposition.
\\

Here is the structure of the paper.  Section \ref{sec:algsum}
is a technical summary
of the algorithm.  Section \ref{sec:algdet}
is the detailed construction.  Section \ref{sec:example}
presents example applications.

\section{Algorithm Summary}\label{sec:algsum}

We compare two sequences $\A, \B$, of lengths $l_\A$ and $l_\B$ 
using a two dimensional array $W$.
The indices of $W$ range from zero to the length of each sequence.

Denote by $\A[i]$ the letter in position $i$ of sequence $\A$,
and by $\A[i_1, i_2]$ the subsequence of $\A$ from positions $i_1$
to $i_2$.

Our metaphor for comparing $\A$ and $\B$ is writing them on
a long paper tape, like CLUSTAL output without line breaks.
Number the columns from zero,
%
%
with the zeroeth position in each row containing a special start symbol,
$\startnode$.  The last column on the tape is filled with a special end symbol,
$\endnode$.  Otherwise, each position may be a letter, A, C, G, T for
nucleotide sequences, or a blank, and no column is entirely blank.
We model the machine which writes the tape column by column as a Markov
process.

A list of transitions for the machine 
is an evolutionary history and has a probability
which is the product of the probability of each transition.

We define the arrays $P(i,j)$, $P^\vee(i,j)$, and $V(i,j)$, with the
same indices as $W$.  Let
$P(i,j)$ be the probability that the machine writes a tape which can
be cut off at some column so that
the first row contains $\startnode$ and $\A[1, i]$
and the second row contains $\startnode$ and $B[1,j]$, interspersed with
any combination of blanks.

This $P$ array produces the \cite{tkf} sum approach when applied with
their family of Markov models. 
Their model gives a sensible way to think of Needleman-Wunsch dynamic
programming summed over all paths, rather than the classic Viterbi 
trace-back path.

Let $P^\vee(i,j)$ be the probability that the machine writes a tape
which can be cut off at some column so that the first row contains
$\A[i+1, l_\A]$ followed by $\endnode$, and the second row 
contains $\B[j+1, l_\B]$ followed by $\endnode$, with any combination
of blanks.
We call $P^\vee$ a back fill array and
$P$ a front fill array.

Let $\Pi_\A$ be the probability for the machine to write a tape with
$\startnode$ followed by $\A$ followed
by $\endnode$ interspersed with blanks in the first row,
and $\Pi_\B$ be the analog for $\B$ in the second row.

Let $$V(i,j) = \log \Pi_\A + \log \Pi_\B - \log P(i,j) - \log P^\vee(i,j).$$

The combination of front fill and back fill
makes the values in array $V$ comparable with each
other as log likelihoods for the evolutionary process
to pass through intermediate points given by coordinates
$(i,j)$.  The normalization by $\Pi_\A$ and $\Pi_\B$
makes $V$ arrays for different sequences comparable.
That is, we can answer whether it is
more likely $\A_1$ and $\B_1$ are related
than $\A_2$ and $\B_2$ are related.

In the detailed algorithm construction, we modify the
TKF process to produce an array $W$.  The concept for
the algorithm is the same, but the array $W(i,j)$ 
weights likelihoods for alignments near position $i$ in $\A$ and $j$ in
$\B$ more heavily than positions far away.

\section{Algorithm construction}\label{sec:algdet}

We describe the adaptation of the family of TKF processes
to our framework.  We refer to the underlying Markov process
of \cite{tkf} and \cite{lmsh} as the {\em ground process}.

\subsection{The one state ground process}

The ground process is a simple model of evolution.
It has predictive strength and permits a clean description
of the new algorithm.  The ground process is really a complex
parameter and a different process may be
substituted without essential changes to our algorithm.
For example, we could use the process of \cite{tkf2},
which allows different substitution
models for transitions and transversions.
\\

The ground process evolution model is that beginning with
an ancestor sequence ${\bf C}$, the sequence of each generation
is mostly copied faithfully to the next generation,
and occassionally letters are
substituted, deleted, or inserted.
The frequency of these events is
expressed per letter, per evolutionary distance, i.e. time.

The ground process is reversible.  The probability 
for a particular letter at a certain point to be deleted is 
the same as the probability to insert that particular letter at a that
point.  Substitution of letter $X$ for letter $Y$ has the same
probability as the reverse substitution.

For sequences $\A$ and $\B$, let $$Pr_{t}(\A,\B)$$
be the probability that $\A$ and $\B$ are separated by
evolutionary distance $t$, and so diverged $t/2$ ago.
Let $$Pr_t(\B | \A)$$ be
the probability that $\A$ evolves to $\B$ over evolutionary distance $t$.
In this view, 
$\Pi_\A$ is the equilibrium probability to observe
$\A$ after a long time and $\Pi_\B$ the equilibrium probability to observe
$\B$.

A consequence of reversibility is
$$Pr_{t}(\A, \B) = \Pi_\A Pr_{t}(\B | \A)
= \Pi_\B Pr_{t}(\A | \B).$$
Thus, it is unnecessary to find the hidden ancestor of
$\A$ and $\B$ in order to evaluate their similarity.

We describe the ground process in two parts: the
development of an evolutionary history, and
the specification of the letters which fill
that history.

The start state of the machine writes column zero
{\tiny\begin{tabular}{c}$\startnode$ \\ $\startnode$\end{tabular}},
on the paper tape, expressed as the ordered pair $(\startnode, \startnode)$.
The transition to each state writes a column on the paper tape,
using $\nucnode$ to stand in for a nucleotide to be selected in the next phase.
The available states and their transition outputs are:
homology, $( \nucnode, \nucnode)$;
deletion, $( \nucnode, -)$;
insertion, $ (- , \nucnode)$;
and
termination, $(\endnode, \endnode)$.
A nonhomology event, including mismatch, is deletion followed by insertion.

An evolutionary history is a sequence of transitions beginning at
start and ending at termination.  Define the {\em fate} of a nonblank
symbol in the $\A$ sequence (top or first member of pair)
as the sequence of columns beginning
with that symbol and ending just before the column with the
next nonblank symbol in the $\A$ sequence.  The fate of $\startnode$
is start followed by some number of insertions.  The fate of
$\nucnode$ is either homology or deletion followed by some number of
insertions.  The fate of $\endnode$ is termination.

The parameters for determining the probability of an evolutionary
history according to a ground process are the insertion rate $\lambda$
expressed per base per time, the deletion rate $\mu$, and the
evolutionary distance $t$.  We compute the probabilities using
the limit for large $K$
of a discrete process with insertion rate $\lambda t/K$ per base,
deletion rate $\mu t/K$ per base.  The resulting probabilities
are expressions in $l = \lambda t$ and $m = \mu t$.

Let $$B = \frac{l - le^{l-m} } {m - l e^{l - m} }.$$
Let $$\a = l/m.$$
(This $B$ is $\lambda \beta(t)$ in \cite{tkf}, and $\a$ is $\lambda
\beta(\infty)$.)

The probability for $\startnode$ to be part of an evolutionary
history with $k$ insertions is $$(1-B)B^k.$$

The probability to extend $\A$ by another letter is $\a$.
The probability to transition to end is $1 - \a$.

Given that $\A$ is extended by another letter, we have the
following probabilities.
The probability for homology followed by zero insertions
is $$H = e^{-m} (1-B).$$
The probability for homology followed by $k$ insertions is
$$HB^k.$$

The probability for deletion followed by zero insertions
is $$E = B / \a.$$
The probability for deletion followed by exactly one insertion is
$$N = (1 - e^{-m} - E)(1-B).$$
The probability for deletion followed by $k > 1$ insertions is
$N B^{k-1}.$

For the start and homology events, every successive
insertion is with probability $B$, but for a deletion event,
the probability of the first insertion is different.
The ground process in \cite{tkf} is presented
as a machine with a deletion state that has different transition
probabilities from the other states for this possible first insertion.

The insight of \cite{hwkmw}, extended in \cite{lmsh},
is the description of the ground process
using one main state and multiple transitions, including a ``forbidden
transition'' with negative transition factor to accomplish the same result.
The forbidden transition is from a deletion event with zero insertions
to an insertion event.

The evolutionary history is completed to a sequence alignment
by writing a letter from the alphabet $A, C, G, T$ in place of each
$\nucnode$.

The ground process has parameters $\pi_A, \pi_C, \pi_G, \pi_T$ which 
define a distribution of letters.  The parameter $\sigma$ determines
the subsitution rate per base per time.  As above, the equations
can be expressed in terms of $s = \sigma t$.

The probability for an insertion
or deletion process to produce a letter $X$ is $\pi_X$.
For a homology event, the probability for the ground process to produce
the pair $(X,Y)$
is $\pi_X f(X,Y)$, where
$$f(X,Y) = (1 - e^{s}) \pi_Y + \delta_{X,Y} e^s,$$
and $\delta_{X,Y}$ is one or zero depending on whether $X$ and $Y$
are the same. \\

The computation of the array $P(i,j)$ for sequences
$\A$ and $\B$ is recursive.
The base cases are $P$ with either index negative is zero
and $$P(0,0) = 1 - B.$$
Thereafter,
\begin{eqnarray*}
P(i,j) & = & P(i-1, j) \cdot B \pi_{\A[i]} +
    P(i, j - 1) \cdot B \pi_{\B[j]} + \\
& & P(i-1, j-1) \cdot \alpha  \pi_{\A[i]} ( N \pi_{\B[j]} + 
H f(\A[i], \B[j]) ) - \\
& &  P(i-1, j-1) \cdot B^2 \pi_{\A[i]}\pi_{\B[j]}.
\end{eqnarray*}

The probability to observe $\A$ and $\B$ 
related by some evolutionary history is
$(1 - \a) P(l_\A, l_\B).$

\subsection{Modifying the one state ground process}
We modify the computation of the array $P$ to compute
a related array $Q$.

We consider further implications of the reversibility
of the ground process.  The parameters $l$ and $m$ are
{\em a priori} two degrees of freedom in the ground process.
If the lengths of $\A$ and $\B$
are informative, setting $l$ and $m$ to maximize the
probability to observe sequences of the given lengths
gives a relationship between $l$ and $m$.  We choose $m$ and
derive $l$.

If the sequences $\A$ 
and $\B$ are subsequences of very very long genomes, the
lengths of $\A$ and $\B$ may be artifacts of truncation.
We express this in the ground process by $l = m$, making
insertions and deletions equally likely.  In this case,
extension of $\A$ is a zero information event,
expressed by $\a = 1$.  Other transition probabilities are computed
in terms of 
$$B = l / (1 + l).$$


Gaps at the ends of alignments could be artifacts
of truncation.  We model this by replacing
$B$ by $\a$ at the edges of the $P$ array.
This models extending the sequences in each direction
infinitely with bases selected from distribution $\pi.$

We normalize by dividing by the probability to observe
sequences $\A$ and $\B$ separately given $l,$ $m,$
and $\pi$ and omitting the factor for the inital state
$1 - B$.  We report separately the log likelihood to observe
the given sequences, $$\log \Pi_\A + \log \Pi_\B.$$

Assembling the above insights, we compute

\begin{eqnarray*}
Q(0,0) & = & 1, \\
Q(i,0) & = & Q(i-1, 0), \\
Q(i,j) & = & Q(i-1, j) \cdot E + Q( i, j-1) \cdot E +\\
& & Q(i-1, j-1) (N + H f(\A[i], \B[j]) / \pi_{B[j]}) -\\
& & Q( i-1, j-1) \cdot E ^2, \\
Q(i,l_{\B}) & = & Q(i-1, l_{\B})  + Q(i, l_{\B} -1) \cdot E + \\
& & Q(i-1, l_{\B} -1) (N + H f(\A[i], \B[ l_\B] / \pi_{B[l_\B]}) - \\
& & Q( i-1, l_{\B}-1) \cdot E^2.
\end{eqnarray*}
We do not multiply the final entry
by $1 - \a$, because we are not asserting
the sequences end.

The $Q$ array provides the same kind of information as the $P$.
The essential difference is that the gap cost for leading and
trailing gaps is canceled.  It is possible to treat gaps
differently along each edge of the array.

\subsection{New constructions for visualization}

Computing evolutionary history using the sequences $\A$ reversed
and $\B$ reversed is computationally the same problem as for
$\A$ and $\B$ as given.
%

Let $P^\vee(i,j)$ be the probability for the sequences $\A[i+1..l_\A]$
and $\B[j+1, l_\B]$ to align by
some evolutionary history.  Let
$Q^\vee(i,j)$ be $P^\vee(i,j)$ normalized
by the probability to observe the given sequences.  

The base cases are \begin{eqnarray*}
P^\vee(l_\A, l_\B) & = & 1, \\
Q^\vee(l_\A, l_\B) & = & 1.
\end{eqnarray*}
 The equations for
$P^\vee(i,j)$ and $Q^\vee(i,j)$ 
are the same as those for $P$ and $Q$, except progress
is towards $i=0, j =0$ rather than away.

The product $P(i,j)P^\vee(i,j)$ is the
probability $\A$ and $\B$ arose from some evolutionary history
which restricts to an evolutionary history for $\A[1, i]$
and $\B[1, j]$, and 
for $\A[i+1, l_\A]$ and $\B[j+1, l_\B]$

Let $$R(i,j) = Q(i,j)Q^\vee(i,j).$$  The entry $R(i,j)$ is the
probability for an evolutionary history as for $P(i,j)P^\vee(i,j)$,
providing for the artifactual truncation of sequence, divided by
the probability that the sequences $\A$ and $\B$ were observed and
the evolutionary history is start, $l_\A$ deletions followed
by $l_\B$ insertions, end.

Every entry in $R(i,j)$ is comparable to every other entry.  The
contours near the maximum value of $R(i,j)$ bound the region containing
the paths of the most likely alignments.
\\


We propose another pair of arrays $S(i,j)$ and $S^\vee(i,j)$ which
weight recent history more than distant history.  We adapt the
standard technique of approximating a sliding window by multiplying
an accumulator by a factor between zero and one, and adding the new
data value.  An alternative interpretation is that we view the
developing evolutionary history through a fog which makes distant
states uncertain.  The thickness of the historical fog is a parameter
$\rho.$  The generic computation for $S$ is
\begin{eqnarray*}
S(i,j) & = & \rho + (1-\rho)S(i-1, j) \cdot E + \\
& & (1-\rho)S( i, j-1) \cdot E)+\\
& & (1-\rho)S(i-1, j-1) (N + H f(\A[i], \B[j]) / \pi_{B[j]}) -\\
& & (1-\rho)^2 S( i-1, j-1) \cdot E ^2, \\
\end{eqnarray*}
with suitable modifications at the boundaries.

Another model with the same mathematical description is
that with probability
$\rho \Pi_{\A[1, i]} \Pi_{\B[1,j]}$
the subsequences $\A[1, i]$ and $\B[1,j]$ are observed
and discarded from consideration in the evolutionary history.
The Smith-Waterman algorithm is a Viterbi path application
of the similar idea that the alignment process might have
optimal score applied to subsequences of $\A$ and $\B$.

We use the parameter $$r = \rho^{-1} -1.$$
From the view of $\rho$ as the probability of a jump event,
we call $r$ the odds against a jump event.

When $r = \infty$, and the other parameters are equal,
$$Q(i,j) = S(i,j).$$

We implement the computations
\begin{eqnarray*}
J(i,j) & = & - \log S(i,j), \\
J^\vee(i,j) & =& -\log S^\vee(i,j), \\
W(i,j) & = & J(i,j) + J^\vee(i,j).
\end{eqnarray*}

We also compute the equilibriuim value $\nu$ for
noise given parameters $l, m, r, s$ and distribution $\pi$.
This is the limit for large $i, j$ of $J(i,j)$ for
long random sequences constructed from distribution $\pi.$

Using $\nu$, we can approximate $W(i,j)$ in any shape 
region by clamping the boundary values of $J(i,j)$
and $J^\vee(i,j)$ equal to $\nu$ when they are not otherwise
actually calculated.  This makes sense, for example, when
we want to avoid computing on an area where we believe there
is no alignment and want results for an adjacent area.
We can also mask out a high scoring evolutionary history
by forcing $J(i,j) = \nu$ on its path.  This allows a second
place ridge to be seen as maximum.

\section{Application of the $W$ array}\label{sec:example}

Computing $W$ requires the sequences $\A$ and $\B$, the
parameters insertion rate $l$, deletion rate $m$,
substitution rate $s$ and
jump odds $r$, and the distribution $\pi$.

In the examples below, we fix $l = m$, and take $\pi$ to
be the distribution observed in the sequences $\A$ and $\B$.
We handle the nucleotide symbol $N$ as a match for each
nucleotide $A, C, G, T$ with distribution $\pi$.

In principle, these parameters can be dynamically reestimated
for different points in the array.  Also, a global maximum
likelihood reestimation can be done in the manner of the
\cite{tkf} sum approach.

We compute, but do not apply, the noise value $\nu$.

We apply a simple digital image technique to find the
local extreme contours.  We compute $W(i,j)$ and plot,
on a grayscale, the difference
$$\Delta W(i,j) = W(i+1,j) - W(i, j+1).$$

In Section \ref{sec:cyt},
we show an example of finding evolutionary distance by using
$r = \infty$ and simple reestimation of $s$.  We show an
example of curating sequences wih repeats and duplications.

Note in the following examples the black and white diamonds
in the plots of $\Delta W$.  These occur at places where
segments of the two sequences, represented on the horizontal and vertical
axes, align with high identity.  The sharp contrast along the
diagonal of the diamond indicates a local extreme contour in the $W$
array.
%
%
The intensity of a sequence identity feature in the $W$ 
array is proportional to its length.  The longest identity 
feature has maximum intensity and depresses the intensity of
other identity features.
In the $\Delta W$ array, the width of an identity feature
is proportional to its length, and its intensity is not
affected by its length.  Thus, $\Delta W$ shows different
length identity features simultaneously.

\subsection{Comparison of three rodent cytochromes}
\label{sec:cyt}

We compare the mRNA sequences for proteins {\it Rattus norvegicus}
cytochrome P450 IIA1 and IIA2, (Cyp2a1 and Cyp2a2), from \cite{rncyp2a},
and {\it Mus musculus} cytochrome P450,
IIA12 (Cyp2a12), from \cite{mmcyp2a12}. 

\begin{figure}[htbp]
\centerline{
\includegraphics{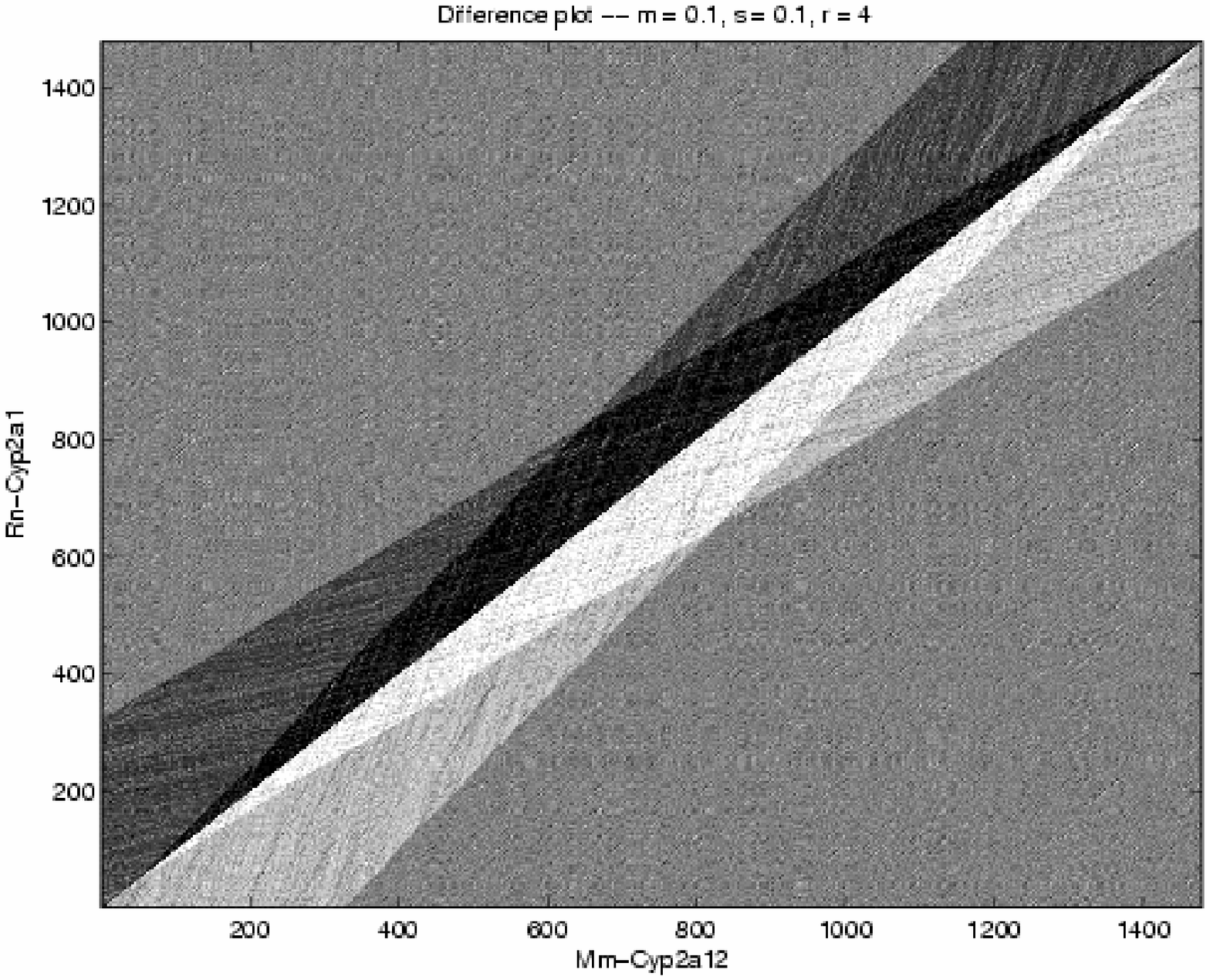}
}
\caption{Comparison of cytochromes Rn-Cyp2a1 and Mm-Cyp2a12.  We
show the log likelihood difference $\Delta W$ in grayscale.  Qualitatively,
the other pairwise comparisons, with Rn-Cyp2a2, look the same.  The unique
sharp diagonal boundary between black and white triangles shows the best
alignment, which far surpasses any alternative model. }
\label{fig:ca1apic}
\end{figure}

Figure \ref{fig:ca1apic} shows a grayscale plot $\Delta W(i,j)$ for
sequences Rn-Cyp2a1 and Mm-Cyp2a12 using the parameters m =
s = 0.1 and r = 4.
From this picture, we can infer that the evolutionary history between these
sequences contains no jump events.  Thus, we set $r = \infty$ for further
iterations. We also see no deletion events, so we decrease $m.$ 

\begin{figure}[htbp]
\centerline{
\includegraphics{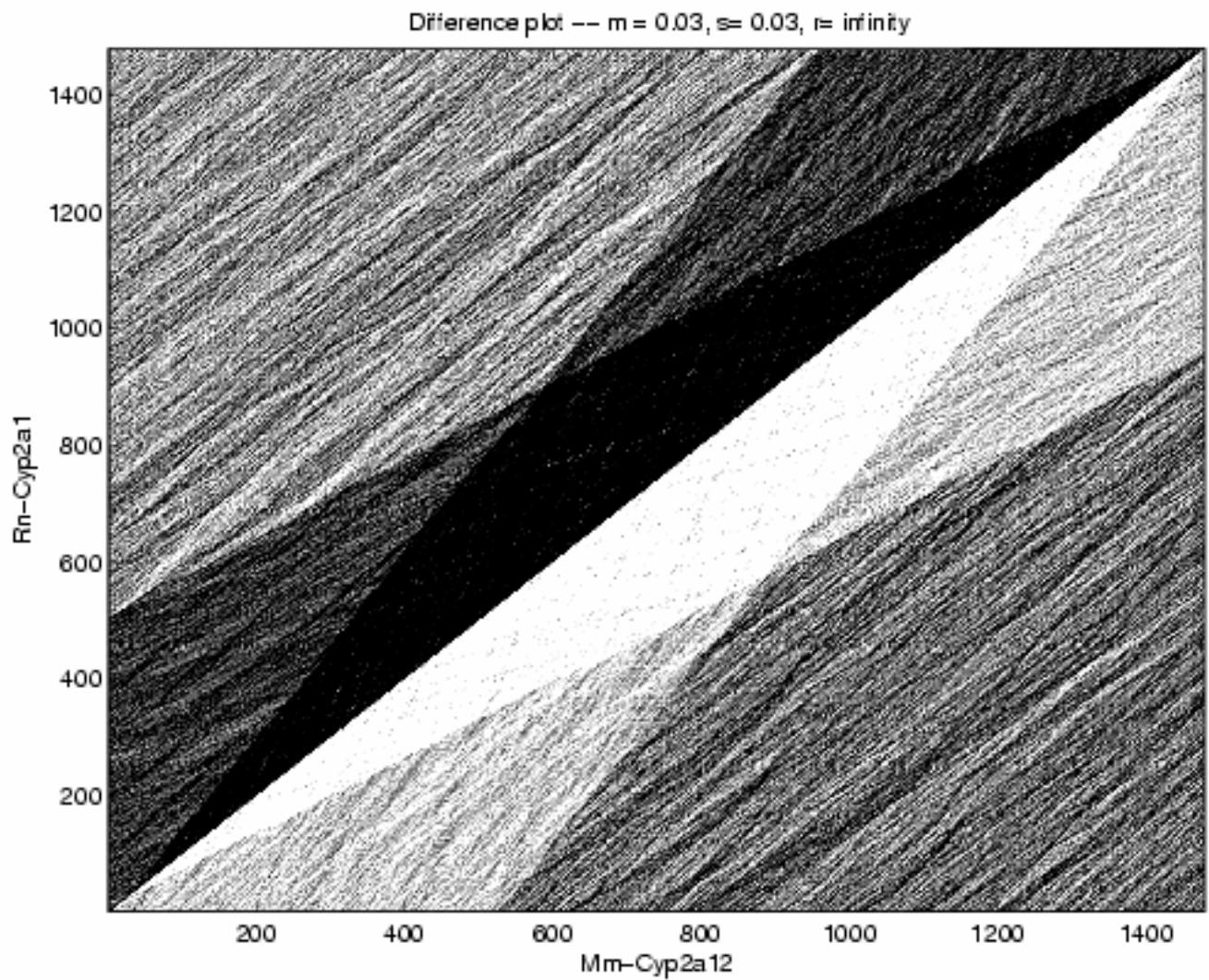}
}
\caption{Comparison of cytochromes Rn-Cyp2a1 and Mm-Cyp2a12, with
parameters derived from the no jump, no deletion hypothesis, $r = \infty$,
decreased $m$ from figure \ref{fig:ca1apic}.  We show the log likelihood
difference $\Delta W$ in grayscale.
}
\label{fig:ca1b}
\end{figure}

Figure \ref{fig:ca1b} shows a grayscale plot of $\Delta W(i,j)$ with
parameters $m = s = 0.03$, $r = \infty$.  We perform this computation
for all three pairs of sequences.  For the following pairs, we
calculated the difference between the maximum and minimum of the $W$
array.
\begin{center}
\begin{tabular}{llr}
Mm-Cyp2a12 & Rn-Cyp2a1 & 2263 \\
Mm-Cyp2a12 & Rn-Cyp2a2 & 2110 \\
Rn-Cyp2a1 & Rn-Cyp2a2 & 2398 \\
Mm-Cyp2a12 & Mm-Cyp2a12 & 2844
\end{tabular}
\end{center}

We expect two random sequences to be about 27\% identical.
The value is greater than 25\%, because the distribution of nucleotides
$\pi$ is not flat.  We know Mm-Cyp2a12 is 100\% identical to
itself.  By dividing 73\% into 2844, we find that
a difference of 38 units of log-likelihood represents
1\% of sequence identity for these sequences and parameters.

We estimate the following sequence divergences:
\begin{center}
\begin{tabular}{llr}
Mm-Cyp2a12 & Rn-Cyp2a1 & 14.9\% \\
Mm-Cyp2a12 & Rn-Cyp2a2 & 18.8\% \\
Rn-Cyp2a1 & Rn-Cyp2a2 & 11.4\% 
\end{tabular}
\end{center}

We compute arrays with new parameters.  We maintain $r = \infty$.
We set $m = 10^{-6}$.  We set $s$ to be the whole number percentages
near the values computed above.  The plots of $\Delta W$ are similar
to those for figures \ref{fig:ca1apic} and \ref{fig:ca1b}.
The values of $s$ which produce the maximum log-likelihoods are:
\begin{center}
\begin{tabular}{llr}
Mm-Cyp2a12 & Rn-Cyp2a1 & 14\% \\
Mm-Cyp2a12 & Rn-Cyp2a2 & 17\% \\
Rn-Cyp2a1 & Rn-Cyp2a2 & 11\% 
\end{tabular}
\end{center}


\subsection{Comparison of two zinc finger proteins}

We compare two zinc finger proteins from {\it Mus musculus}.
We use the mRNA sequences for Zfp111 and Zfp235 \cite{zfp}.
These two proteins have a similar
amino acid sequences: one KRAB domain, followed by a spacer region,
followed by a series of zinc finger domains in tandem.  The members of
this protein family each have from five to nineteen of the
28 amino acid zinc finger domains.
Most of these amino acids are required
for zinc binding, and are highly conserved between duplications and
between genes.  Shannon et al. show statistically that
over all nucleotide positions in the zinc finger domains,
substitution events tend to be synonymous.  They observe
a range of selection behaviors at positions
believed to be noncritical to the zinc binding function.

\begin{figure}[htbp]
\centerline{
\includegraphics{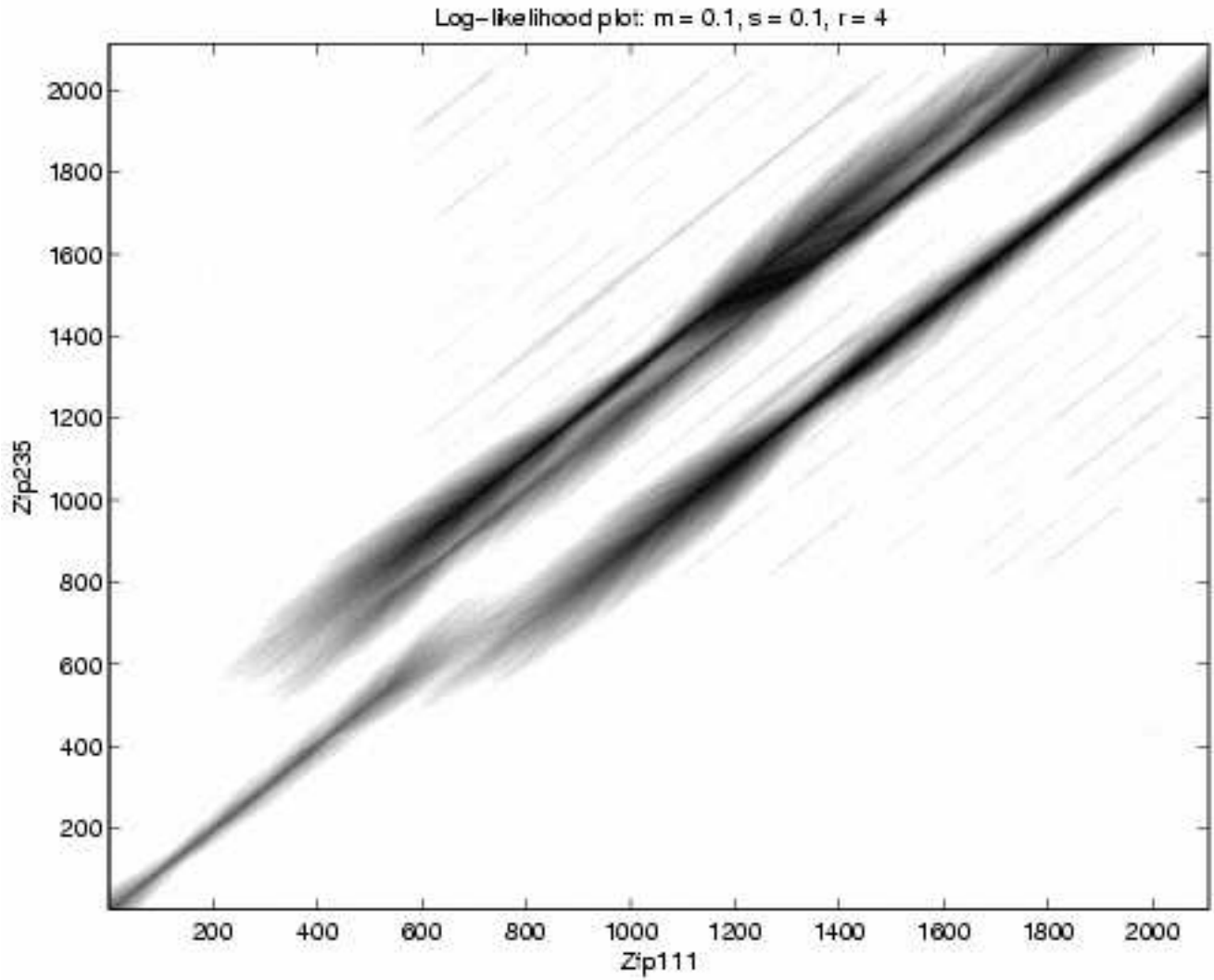}
}
\caption{Comparison of zinc finger protein mRNA seqences for
{\it Mus musculus} Zfp111 and Zfp235.  The normalized likelihood array $W$
shows: high sequence identity to coordinate (537, 540), then a mismatch
with a net insertion of 222 bases in Zfp235.  In the rectangle
with bottom left corner (622, 844) and top right corner (2106, 2112),
there are many parallel tracks, with parts of three more intense than the
others.
}
\label{fig:gbza2}
\end{figure}

Figure \ref{fig:gbza2} shows the $W$ array computed
with the Zfp111 and Zfp235 sequences.  The $W$ array shows high sequence
identity at the beginning of the sequence, followed by
a mismatch with a net insertion in the Zfp235 sequence.

\begin{figure}[htbp]
\centerline{
\includegraphics{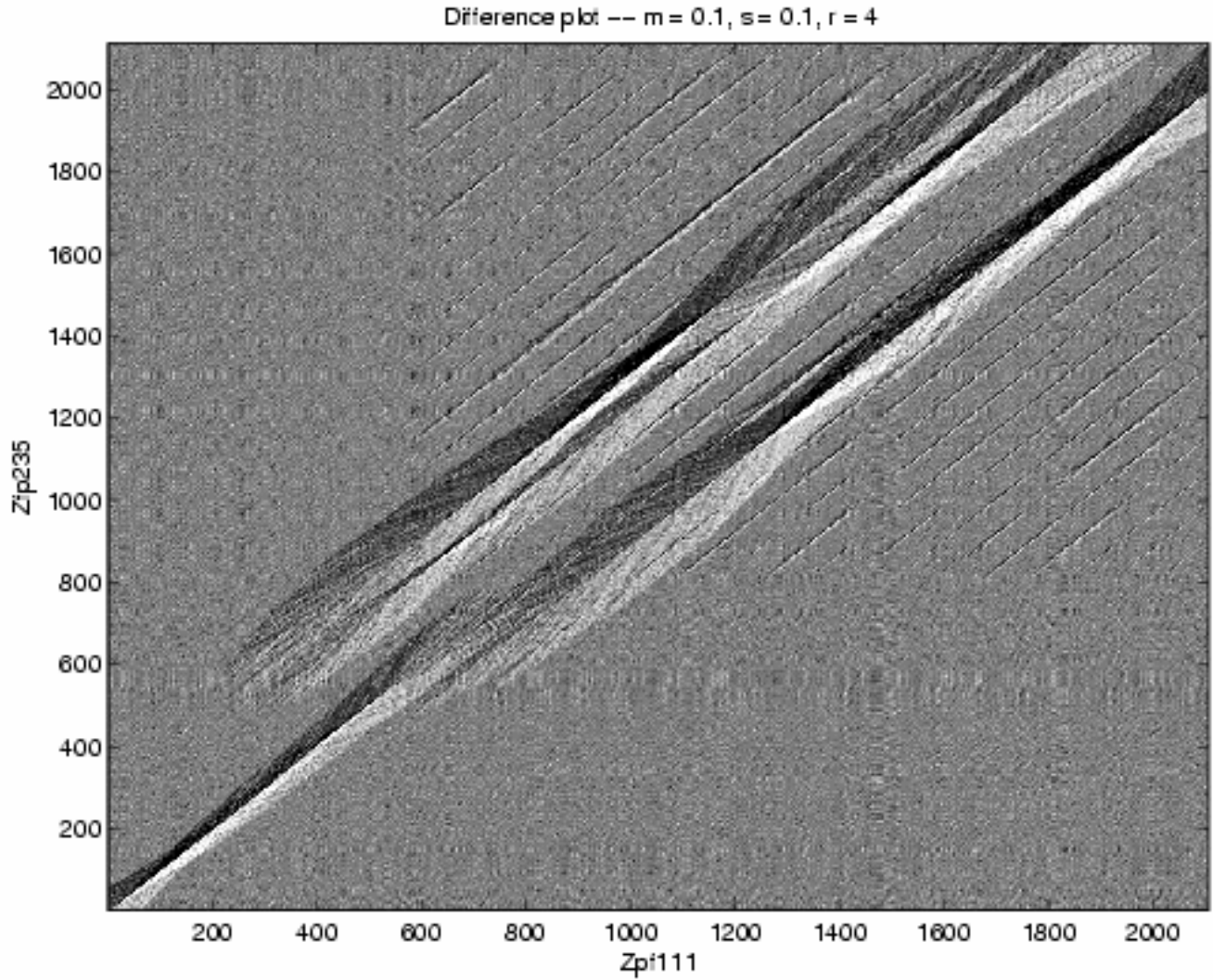}
}
\caption{Difference plot $\Delta W$ from comparison of Zfp111 and Zfp235.
The $\Delta W$ array suggests there is high sequence identity between
bases 547 through 621 of Zfp111 and 722 through 843 of Zfp235 relative
to any potential alignment of these two sequences, by the narrow black
and white diamond at these coordinates.
Above and to the right of coordinate (622, 844), 
the parallel
tracks suggest near-repeat subsequences and the black
and white diamonds indicate the more similar regions.
The diamonds which cover the same vertical coordinates,
near coordinates 1300 and 1800 of Zfp235,
suggest sequence duplication within Zfp111.
}
\label{fig:gbda}
\end{figure}

Figure \ref{fig:gbda} shows the $\Delta W$ plot for
the same sequences and parameters.  The $\Delta W$ plot
provides more detail on the mismatch discovered by the $W$
array plot.  We see insertion of the second Zfp235 zinc finger
relative to Zfp111 and insertion of either the
eighth or ninth Zfp111 zinc finger relative to Zfp235.

\begin{figure}[htbp]
\centerline{
\includegraphics{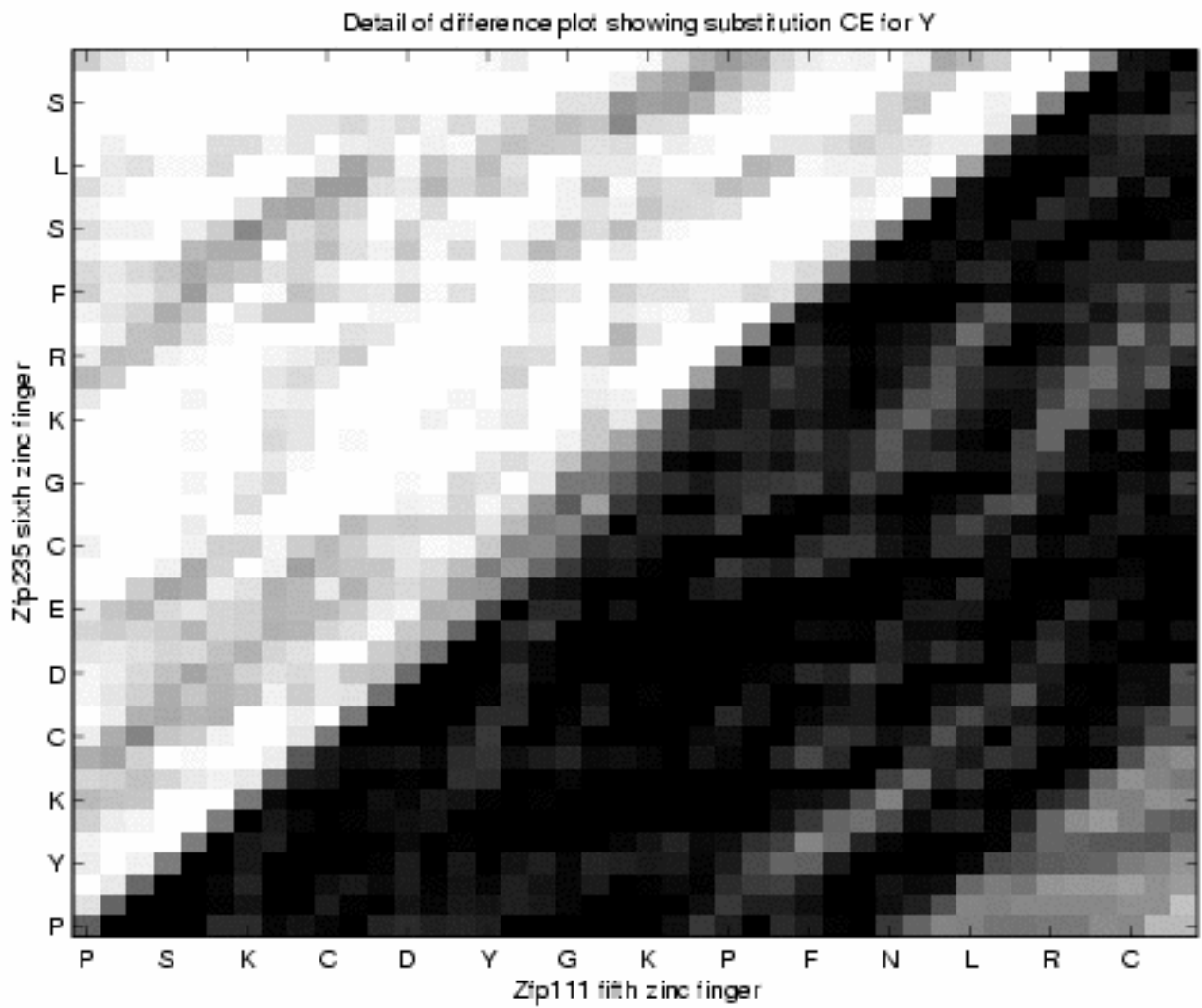}
}
\caption{Detail of $\Delta W$ showing single amino acid deletion of
either  the sixth or seventh amino acid of the fifth zinc finger in Zfp111.}
\label{fig:gbda56}
\end{figure}

Figure \ref{fig:gbda56} shows a detail of the same $\Delta W$
array.  We can locate a single amino acid deletion in Zfp111
relative to Zfp235.

\begin{figure}[htbp]
\centerline{
\includegraphics{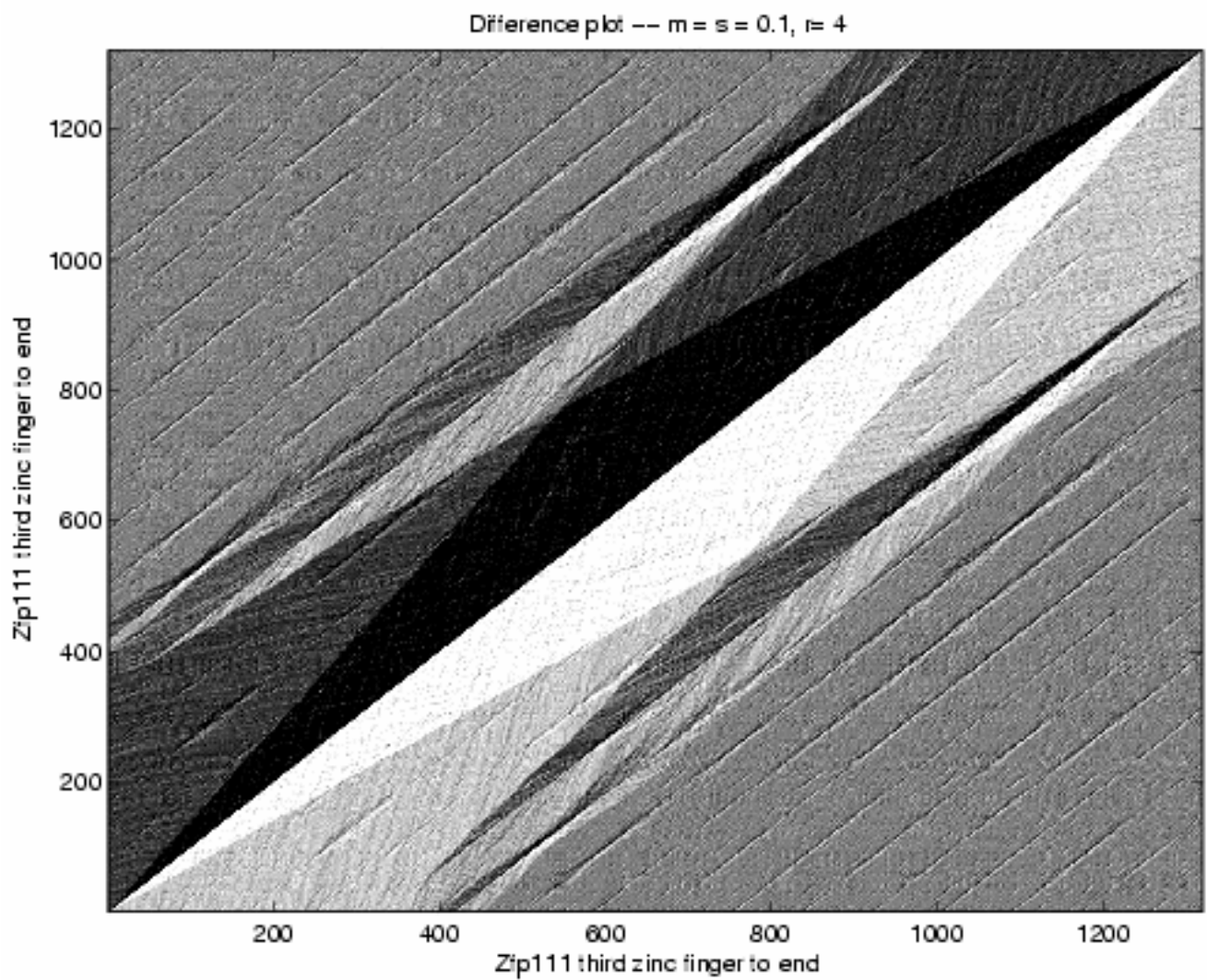}
}
\caption{The $\Delta W$ array of Zfp111 compared with itself.
The black and white diamonds off the large one, centered near
coordinates $(700,1100)$ and $(1100, 700)$ strongly suggest
internal duplication.
}
\label{fig:gbdd}
\end{figure}

Zinc fingers 10 through 13 of Zfp235 match zinc
fingers 10 through 13 and also zinc fingers 14 through 17
of Zfp111.  Figure \ref{fig:gbdd}, showing the $\Delta W$
array for Zfp111 against itself, provides strong evidence
of the internal duplication in Zfp111.

\section{Acknowledgment}
We thank Rick Durrett for suggesting the examples.

\noindent
Department of Mathematics \\
Malott Hall \\
Cornell University \\
Ithaca, NY 14853 \\
USA \\
\\
{\tt lawren@math.cornell.edu}

\end{document}